\documentclass[reprint,
superscriptaddress,
 amsmath, amssymb,
 aps, times,
 pra,
]{revtex4-2}

\usepackage{graphicx}% Include figure files
\usepackage{dcolumn}% Align table columns on decimal point
\usepackage{bm}% bold math

\usepackage[colorlinks=true,urlcolor=blue,citecolor=blue,linkcolor=blue]{hyperref}%  
\usepackage{xcolor}
\usepackage[braket, qm]{qcircuit}
\usepackage{mathtools} % for bsmallmatrix
\usepackage{upgreek} %  for upmu
\usepackage{bbm}
\usepackage{wasysym} % for \diameter

\usepackage{enumerate}

\usepackage{ragged2e}
\usepackage{algorithm}
\usepackage{algpseudocode}

\newcounter{mycounter} 
\newcommand{\FF}{\mathbb{F}}
\newcommand{\RR}{\mathbb{R}}

\newcommand{\ZZ}{\mathbb{Z}}

\newcommand{\GL}{\text{GL}}
\newcommand{\diag}{\text{diag}}
\newcommand{\CZ}{\textsc{\MakeLowercase{CZ}}}

\newcommand{\Tr}{\text{Tr}}
\newcommand{\rank}{\text{rank}}
\newcommand{\poly}{\text{poly}}

\newcommand{\unitspace}{\mskip3mu}

\definecolor{mygray}{HTML}{eaefef}
\definecolor{mygray2}{HTML}{f0ffff}
\definecolor{mygreen}{HTML}{7BCC70}
\definecolor{mygreen2}{HTML}{79df6c}
\definecolor{myblue2}{HTML}{87CEFA}
\definecolor{myblue}{HTML}{5f86ff}
\definecolor{myyellow}{HTML}{fed321}
\definecolor{myred}{HTML}{fd7062}

\definecolor{darkred}{HTML}{900000}
\definecolor{darkgreen}{HTML}{006000}
\definecolor{darkblue}{HTML}{000090}
\definecolor{darkyellow}{HTML}{c08000}

\newcommand{\crossrefcolor}{\color{red}}
\renewcommand{\crossrefcolor}{}

\newcommand{\solver}[1]{\texttt{#1}}

\newcommand{\fu}{Dahlem Center for Complex Quantum Systems, Freie Universit{\"a}t Berlin, 14195 Berlin, Germany}
\newcommand{\fzj}{Institute for Theoretical Nanoelectronics (PGI-2), Forschungszentrum J\"ulich, 52428 J\"ulich, Germany}
\newcommand{\ibm}{IBM Quantum, IBM Research Europe – Zurich, S\"aumerstrasse 4, 8803 R\"uschlikon, Switzerland}
\newcommand{\pasqal}{Current address: PASQAL, 7 rue Léonard de Vinci, 91300 Massy, France}

\begin{document}

\preprint{APS/123-QED}

\title{Hardware-Tailored Diagonalization Circuits}
\author{Daniel Miller} 
\email{daniel.miller@fu-berlin.de}

\affiliation{\fu}
\affiliation{\fzj}
\affiliation{\ibm}

\author{Laurin E. Fischer}
\affiliation{\ibm}

\author{Kyano Levi}
\affiliation{\fu}

\author{Eric J. Kuehnke}
\affiliation{\fu}

\author{\\Igor O. Sokolov}
\affiliation{\ibm}
\affiliation{\pasqal}

\author{Panagiotis Kl. Barkoutsos}
\affiliation{\ibm}
\affiliation{\pasqal}

\author{{Jens Eisert}}
\affiliation{\fu}

\author{Ivano Tavernelli}
\affiliation{\ibm}

\begin{abstract}
A central building block of many quantum algorithms is the diagonalization of Pauli operators.
Although it is always possible to construct a quantum circuit that simultaneously diagonalizes a given set of commuting Pauli operators,
only resource-efficient circuits can be executed reliably on near-term quantum computers.
Generic diagonalization circuits, in contrast, often lead to an unaffordable \textsc{Swap} gate overhead on quantum devices with limited hardware connectivity.
A common alternative is to exclude two-qubit gates altogether.
However, this comes at the severe cost of restricting the class of diagonalizable sets of Pauli operators to tensor product bases (TPBs).
In this article, we introduce a theoretical framework for constructing hardware-tailored (HT) diagonalization circuits.
Our framework establishes a systematic and highly flexible procedure for tailoring diagonalization circuits with ultra-low gate counts.
We highlight promising use cases of our framework and 
-- as a proof-of-principle application -- 
we devise an efficient algorithm for grouping the Pauli operators of a given Hamiltonian into jointly-HT-diagonalizable sets.
For several classes of Hamiltonians, we observe that our approach requires fewer measurements than conventional TPB approaches. 
Finally, we experimentally demonstrate that  
HT circuits can improve the efficiency of estimating expectation values with cloud-based quantum computers. 

\end{abstract}

\maketitle

\date{\today}

\section*{Introduction}

Since first-generation quantum computers were made publicly available eight years ago by IBM~\footnote{IBM, the IBM logo, and ibm.com are trademarks of International Business Machines Corp., registered in many jurisdictions worldwide. Other product and service names might be trademarks of IBM or
other companies. The current list of IBM trademarks is available at \url{https://www.ibm.com/legal/copytrade}.}, the technological frontier is expanding at an ever increasing pace~\cite{kandala_hardware_efficient_2017, arute_quantum_supremacy_2019, krantz_a_quantum_2019,  blais_circuit_quantum_2021, bravyi_the_future_2022, burkard_semiconductor_spin_2023,  krinner_realizing_repeated_2022, scalable_error_kandala_2023, kim_evidence_for_2023, hangleiter_computational_advantage_2023, bluvstein_logical_quantum_2024}. 
Nevertheless, decoherence and hardware errors still limit the applicability of these early-stage quantum devices, and practical quantum advantage yet remains to be demonstrated.
To go beyond what is possible now, it is crucial to operate both classical and quantum com\-pu\-ters in an orchestrated manner that exploits their respective strengths.

For example, in the \emph{variational quantum eigensolver} (VQE) algorithm~\cite{peruzzo_a_variational_2014, McClean_the_theory_2016, mcardle_quantum_computational_2020}, a classical computer optimizes the parameters of a trial quantum state vector $\ket{\psi}$ to accurately approximate the minimal eigenvalue of an observable $O$, e.g., the Hamiltonian of a molecule.
The tasks performed by the quantum processor are preparing $\ket{\psi}$ and 
gathering measurement data from which the expectation value $\langle{O}\rangle=\bra{\psi}O\ket{\psi}$ can be estimated. 
In practice,  the observable $O$ cannot be measured directly as this would require a quantum circuit that diagonalizes it, i.e., a circuit that rotates the unknown eigenbasis of $O$ to the computational basis. 
A common approach to circumvent this problem is to express $O$ as a linear combination of $n$-qubit Pauli operators $P_i\in\{I,X,Y,Z\}^{\otimes n}$ with real coefficients $c_i\in \RR$, as in
\begin{equation}\label{eq:observable_decomposition}
O = \sum_{i=1}^{M} c_i P_i.
\end{equation}
Since any Pauli operator $P_i$ can be diagonalized using single-qubit Clifford gates, it is straightforward to measure its expectation value $\langle P_i \rangle$. 
Once all $\langle P_i \rangle$ have been obtained, $\langle{O}\rangle$ is calculated from Eq.~\eqref{eq:observable_decomposition}.
Although the number $M$ of Pauli operators for molecular Hamiltonians has a nominal scaling of up to $\mathcal{O}(n^4)$,
measuring all Pauli expectation values individually would require a large number of quantum circuit executions (``shots'')~\cite{mcardle_quantum_computational_2020}.
In addition to the mere evaluation of $\langle O \rangle$, usually  also its gradient is  estimated from the measured data \cite{schuld_evaluating_analytic_2019, sweke_stochastic_gradient_2020}.
To keep resource requirements at an affordable level,
one can make use of simultaneous measurements of commuting Pauli operators,
a technique that is commonly applied.
For example, with a diagonalization circuit that only contains single-qubit gates,
one can measure a \emph{tensor product basis} (TPB), i.e., a set of Pauli operators that are \emph{qubitwise commuting} (QWC)~\cite{kandala_hardware_efficient_2017}.
For typical problems it is possible to group an average number of three Pauli operators into a common TPB~\cite{verteletskyi_measurement_optimization_2020}.
To further reduce the number of required diagonalization circuits from $\mathcal{O}(n^4)$ to $\mathcal{O}(n^3)$, grouping the Pauli operators into \emph{general commuting} (GC) sets 
has been suggested~\cite{gokhale_ON3_Measurement_2020, yen_measuring_all_2020, jena_pauli_partitioning_2019}.
Under ideal circumstances, such GC groupings would substantially decrease the number of shots required to estimate $\langle O \rangle$ to a desired accuracy~\cite{crawford_efficient_quantum_2021}.
Unfortunately, the corresponding diagonalization circuits consist of up to $\tfrac{n(n-1)}{2}$ two-qubit gates and, if the connectivity of the device is limited, a large number of additional \textsc{Swap} gates.
As a first step to interpolate between these extremes, diagonalization circuits featuring a single layer of two-qubit gates have been introduced~\cite{hamamura_efficient_evaluation_2019, escudero_hardware_efficient_2022}; however, it is an open challenge to fully exploit the trade-off between QWC and GC.
As of today, the best method to experimentally estimate an expectation value $\langle O \rangle$ is unclear.
Besides Pauli grouping~\cite{kandala_hardware_efficient_2017, jena_pauli_partitioning_2019, hamamura_efficient_evaluation_2019, verteletskyi_measurement_optimization_2020, gokhale_ON3_Measurement_2020, yen_measuring_all_2020, crawford_efficient_quantum_2021, wu_overlapped_grouping_2021, yen_deterministic_improvements_2023, escudero_hardware_efficient_2022}, 
there is active research 
in addressing this problem with classical shadows~\cite{ohliger_efficient_and_2013, aaronson_shadow_tomography_2018, huang_predicting_many_2020, hadfield_measurements_of_2020, huang_efficient_estimation_2021,  hadfield_adaptive_pauli_2021}, 
unitary partitioning~\cite{izmaylov_unitary_partitioning_2020, zhao_measurement_reduction_2020}, 
low-rank factorization~\cite{huggins_efficient_and_2021},
adaptive estimators~\cite{garcia_perez_learning_to_2021, fischer_ancilla_free_2022, shlisberg_adaptive_estimation_2023}, and
decision diagrams~\cite{hillmich_decision_diagrams_2021}.

In this article, we introduce a theoretical framework for constructing diagonalization circuits whose two-qubit gates are tailored to meet the connectivity restrictions imposed by most current quantum computing architectures, e.g., super- and semiconducting qubits~\cite{kandala_hardware_efficient_2017, arute_quantum_supremacy_2019, krantz_a_quantum_2019,  blais_circuit_quantum_2021, bravyi_the_future_2022, krinner_realizing_repeated_2022, burkard_semiconductor_spin_2023,  scalable_error_kandala_2023, kim_evidence_for_2023}. 
Our flexible approach can be applied to any hardware connectivity. 
We demonstrate the viability of our techniques for a large class of paradigmatic Hamiltonians in the context of the Pauli grouping problem.

\section*{Results}

\subsection*{Framework for HT Diagonalization Circuits}
The purpose of our theoretical framework is the construction of \emph{hardware-tailored} (HT) Clifford circuits that diagonalize a given set of commuting $n$-qubit Pauli operators $P_1,\ldots,P_m$.
After possibly replacing some of the $P_j$ by $-P_j$, we can assume that the group they generate, 
which is denoted by $\langle P_1,\ldots, P_m \rangle$, 
does not contain $-I^{\otimes n}$.
From now on, we will always assume that this is the case since it allows us
to extend  $\langle P_1,\ldots, P_m \rangle$ to the {stabilizer group} $\mathcal{S}$
of some stabilizer state vector $\ket{\psi_\mathcal{S}}$, where $\ket{\psi_\mathcal{S}}$ is defined as the common $+1$-eigenvector of all operators $S\in \mathcal{S}$, see  the \emph{Supplementary Material} {\crossrefcolor (SM) Sec.~III}.
%{\blue(methods)}~\cite{gottesman_phd_thesis_1997,dehaene_clifford_group_2003}.
Then, uncomputing the state vector $\ket{\psi_\mathcal{S}}$, i.e., applying some Clifford circuit $U_\mathcal{S}^\dagger$ for which $\ket{\psi_\mathcal{S}} = U_\mathcal{S}\ket{0}^{\otimes n}$, will simultaneously diagonalize $P_1,\ldots, P_m$~\cite{dehaene_clifford_group_2003}.

An important class of stabilizer states is  that of graph states~\cite{hein_multiparty_entanglement_2004}. 
A graph with $n$ vertices is defined in terms of its
{adjacency matrix} $\Gamma = (\gamma_{i,j})\in \FF_2^{n\times n}$, 
where $\FF_2$ is the binary field;
a pair $(i,j)$ of vertices is connected via an edge if and only if $\gamma_{i,j}=1$.
In this article, we do not distinguish between a graph and its adjacency matrix, and we follow the convention $\gamma_{i,j}=\gamma_{j,i}$ and $\gamma_{i,i}=0$ for all $i,j\in\{1,\ldots,n\}$.
Every graph $\Gamma$ defines a graph state vector $\ket{\Gamma} = U_\Gamma \ket{0}^{\otimes n}$ whose preparation circuit 
\begin{equation} \label{eq:circ_graph}
U_\Gamma = \Bigg( \prod_{i<j} \CZ ^{\gamma_{i,j}}_{i,j} \Bigg)  H^{\otimes n}
\end{equation}
consists of a layer of Hadamard gates $H=\tfrac{1}{\sqrt{2}}(X+Z) $, followed by  a two-qubit gate $\CZ= \diag(1,1,1,-1)$  for every pair of connected vertices.
The stabilizer group of $\ket{\Gamma}$ is  $\mathcal{S}_\Gamma = \{  X^\mathbf{k} Z^{\Gamma \mathbf{k}}(-1)^{\sum_{i<j}  k_i \gamma_{i,j} k_j } \ \vert \ \mathbf{k} \in \FF_2^n  \}$,
where we define $X^\mathbf{k} = X^{k_1}\otimes\ldots \otimes X^{k_n}$ and similarly  $Z^{\Gamma \mathbf{k}}$ for the matrix-vector product $\Gamma \mathbf{k}$. %~\cite{hein_multiparty_entanglement_2004}.
Hence, $U_\Gamma^\dagger$ would diagonalize our operators $P_1,\ldots, P_m$ if only they were of the form  $\pm X^{\mathbf{k}} Z^{\Gamma \mathbf{k}}$.

\begin{figure}[t]
\centering
\includegraphics[width=\columnwidth]{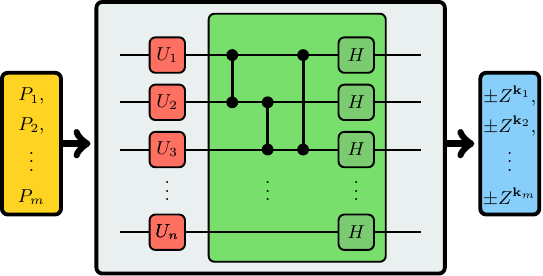}
\caption{\label{fig:diagonalization_circuit}Graph-based diagonalization circuit. A set of commuting Pauli operators $P_1,\ldots, P_m$ (yellow) is diagonalized in two steps:
first, a layer of single-qubit Clifford gates $U= U_1\otimes\ldots \otimes U_n$ (red) 
rotates them into a set of the form $\mathcal{S}=\{\pm X^\mathbf{k} Z^{\Gamma \mathbf{k}} \ \vert \ \mathbf{k}\in \FF_2^n\}$, where $\Gamma\in\FF_2^{n\times n}$ is an adjacency matrix.
Afterward, $\mathcal{S}$ is rotated to the computational basis (blue) by uncomputing the graph state vector $\ket{\Gamma}$ (green).
The existence of $U$ and $\Gamma$ is guaranteed because every stabilizer state is LC-equivalent to a graph state~\cite{vandennest_graphical_description_2004}. 
We call a graph-based diagonalization circuit \emph{hardware-tailored} (HT) if $\Gamma$ is a subgraph of the connectivity graph $\Gamma_\text{con}$ of the considered quantum device.} 
\end{figure}

Every stabilizer state %$\ket{\psi_\mathcal{S}}$ 
is \emph{local-Clifford} (LC) equivalent to a graph state~\cite{vandennest_graphical_description_2004}.
Thus, there exist single-qubit Clifford gates $U_1, \ldots, U_n$ and a graph $\Gamma$ such that $ (U_1\otimes\ldots \otimes U_n) \ket{\psi_\mathcal{S}}=\ket{\Gamma}$.
We conclude that every set of commuting Pauli operators can be simultaneously diagonalized by some layer of single-qubit Clifford gates followed by some circuit of the form $U_\Gamma^\dagger$.
We refer to this procedure as a {graph-based diagonalization circuit}, see Fig.~\ref{fig:diagonalization_circuit}.

The \emph{connectivity graph} of a quantum computer
is the graph $\Gamma_\mathrm{con}$ whose vertices and edges, respectively, are given by qubits and pairs of qubits for which a $\CZ$~gate can be physically implemented.
For quantum devices with a limited connectivity, 
general graph-based diagonalization circuits require up to $\mathcal{O}(n^2)$ \textsc{Swap} gates~\cite{maslov_linear_depth_2007}.
This overhead renders unconstrained graph-based diagonalization circuits infeasible for certain near-term applications, see
{\crossrefcolor SM Sec.~II}.
However, if $\Gamma\subset \Gamma_\mathrm{con}$ is a subgraph of the connectivity graph, 
\textsc{Swap} gates are avoided completely.
We refer to graph-based diagonalization circuits that are designed to meet this condition as \emph{hardware-tailored} (HT).
By construction, HT circuits have a very low $\CZ$ depth that is upper bounded by the maximum degree of $\Gamma_\text{con}$, e.g., by 3 and 4 for heavy-hex and square-lattice connectivity, respectively~\cite{chamberland_topological_and_2020}.

We now derive a technical condition for the existence of HT diagonalization circuits.
Every $n$-qubit Clifford gate $U$ defines a symplectic matrix
\begin{equation} \label{eq:block_structure}
A = \begin{bmatrix} A^{xx} & A^{xz} \\ A^{zx} & A^{zz} 
\end{bmatrix}
\in \GL(\FF_2 ^{2n})
\end{equation}
with the property that
\begin{equation}\label{eq:trafo1}
UX^\mathbf{r}Z^\mathbf{s} U^\dagger = \text{i}^{\alpha(\mathbf{r},\mathbf{s})} \, X^{A^{xx}\mathbf{r}+ A^{xz} \mathbf{s}} \, Z^{A^{zx}\mathbf{r}+ A^{zz} \mathbf{s}}
\end{equation}
holds for all vectors $\mathbf{r},\mathbf{s}\in \FF_2^n$, see Tab.~\ref{tab:clifford_signs} for the single-qubit case~\cite{dehaene_clifford_group_2003}. 
 \begin{table}
    \caption{Binary representation of the single-qubit Clifford group $\mathcal{C}_1$.
    Every $U\in \mathcal{C}_1$ is a product of $H$ and $S=\diag(1, \text{i})$.
    The six matrices $A\in \GL(\FF_2^2)$ isomorphically correspond to the permutations of $\{X,Y,Z\}$.
    %, reflecting the fact that Pauli operators $X^rZ^s$ are transformed into $UX^rZ^s U^\dagger = i^{\alpha(r,s)}X^{a^{xx}r + a^{xz} s} Z^{a^{zx}r + a^{zz}s}$. 
    }
    \label{tab:clifford_signs}
    \begin{ruledtabular}
     \begin{tabular}{c|cccccc}
         $U$ & $I$ &  $H$ &  $S$ &  $HSH$ & $HS$ &    $SH$  \\\hline 
        $UXU^\dagger$& $X$ & $Z$ & $\text{i}XZ$ & $X$ & $-\text{i}XZ$ & $Z$ \\
        $UZU^\dagger$& $Z$ & $X$ & $Z$ & $-\text{i}XZ$ & $X$ & $\text{i}XZ$  \\ \hline
        $\alpha(0,1)$ & $0$ & $0$ & $0$ & $3$ & $0$ & $1$ \\
        $\alpha(1,0)$ & $0$ & $0$ & $1$ & $0$ & $3$ & $0$ \\
        $A= \begin{bmatrix} a^{xx} & a^{xz} \\
                         a^{zx} & a^{zz} \end{bmatrix}$ &
        $\begin{bmatrix} 1&0\\0&1 \\ \end{bmatrix}$  &
        $\begin{bmatrix} 0&1\\1&0 \\ \end{bmatrix}$  &
        $\begin{bmatrix} 1&0\\1&1 \\ \end{bmatrix}$  &
        $\begin{bmatrix} 1&1\\0&1 \\ \end{bmatrix}$  &
        $\begin{bmatrix} 1&1\\1&0 \\ \end{bmatrix}$  &
        $\begin{bmatrix} 0&1\\1&1 \\ \end{bmatrix}$      \\
     \end{tabular}
 \end{ruledtabular} 
 \end{table}
Hereby, a matrix $A$ is called {symplectic} if $A^\text{T} \begin{bsmallmatrix}0&\mathbbm{1} \\ \mathbbm 1 &0\end{bsmallmatrix} A = \begin{bsmallmatrix}0&\mathbbm{1} \\ \mathbbm 1 &0\end{bsmallmatrix}$, 
with  $\GL(\FF_2^{2n})$ denoting the general linear group of $ \FF_2^{2n}$.
For the time being, we can neglect the phases that are given by $\alpha(\mathbf{r},\mathbf{s}) \in 
\ZZ/4\ZZ=
\{0,1,2,3\}$ 
(see {\crossrefcolor Eq.~(46)} in {\crossrefcolor SM Sec.~XIII} for how to recover them) and focus on the irreducible representation
\begin{equation}
    \mathcal{C}_n \longrightarrow \GL(\FF_2^{2n}), \hspace{8mm} U \longmapsto A
\end{equation}
of the $n$-qubit Clifford group $\mathcal{C}_n$.
If $U = U_1\otimes\ldots \otimes U_n$ is a single-qubit Clifford layer, 
the blocks in Eq.~\eqref{eq:block_structure} are diagonal, i.e., 
$A^{xx} = \diag(a^{xx}_1, \ldots, a^{xx}_n)$,
and similarly for $A^{xz}$, $A^{zx}$, and $A^{zz}$. Hereby,
$a^{xx}_1$ is given by the $xx$-entry of the binary representation of $U_1$, etc.
To construct a single-qubit Clifford layer that rotates $P_1, \ldots, P_m$ into the set $\{\pm X^\mathbf{k} Z^{\Gamma\mathbf{k}}\}$ for a graph $\Gamma$, 
we write $P_j = \text{i}^{q_j}X^{\mathbf{r}_j}Z^{\mathbf{s}_j}$.
By Eq.~\eqref{eq:trafo1}, the application of $U$, which is represented by $A$, transforms the operator $P_j$ into
$P_j' = X^{\mathbf{k}_j} Z^{A^{zx}\mathbf{r}_j+A^{zz}\mathbf{s}_j}$
(up to a global phase), where we have introduced the notation $\mathbf{k}_j=A^{xx}\mathbf{r}_j+A^{xz}\mathbf{s}_j$.
Thus, $P_j' \in \{\pm X^\mathbf{k} Z^{\Gamma\mathbf{k}}\}$ is equivalent to $\Gamma \mathbf{k}_j = A^{zx}\mathbf{r}_j+A^{zz}\mathbf{s}_j$.
These equivalent conditions can be phrased for all $j\in \{1,\ldots, m\}$ simultaneously as a binary matrix equation
\begin{equation} \label{eq:linear_condition_A}
\Gamma A^{xx}R + \Gamma A^{xz}S = A^{zx} R + A^{zz}S,
\end{equation}
where  
$R=\left(\mathbf{r}_1\ \cdots \ \mathbf{r}_m\right)$ and $S=\left(\mathbf{s}_1\ \cdots \ \mathbf{s}_m\right)$ are the two matrices that store the exponent vectors of $P_1, \ldots, P_m$  as their columns.

Equation~\eqref{eq:linear_condition_A} is our first main result.
First, 
by devising algorithms to solve it, we can tackle the challenge of constructing HT diagonalization circuits.
This finally allows us to explore the trade-off between single-qubit Clifford layers and unrestricted Clifford circuits.
On a related note, merely checking whether a guess $(A,\Gamma)$ solves Eq.~\eqref{eq:linear_condition_A} is of course much more efficient than solving Eq.~\eqref{eq:linear_condition_A} from scratch. 
Therefore, our formulation of Eq.~\eqref{eq:linear_condition_A} supports the design of HT diagonalization circuits via educated guesses or dedicated case-based research for, in principle, arbitrarily large system sizes.
We illustrate this idea in Tab.~\ref{tab:guessed_ht_circuits} for molecular Hamiltonians with up to $n=120$ qubits.
 
\begin{table}[]  
    \centering
    \caption{Performance of an educated guess $(A,\Gamma)$ as measured by the number $m$ of jointly-HT-diagonalizable $n$-qubit Pauli operators $P_1,\ldots, P_m $ occurring in a hydrogen chain Hamiltonian $O=\sum_{i=1}^M c_iP_i$ for which $(A,\Gamma)$ solves Eq.~\eqref{eq:linear_condition_A}. The guess corresponds to the constant-depth circuit $H^{\otimes n} (\prod_{k=1}^{n/4} \CZ_{4k+1,4k+2} \CZ_{4k+2,4k+3}) H^{\otimes n} $ which makes use of $n/2$ two-qubit gates and is tailored to a linear hardware connectivity.
    To diagonalize the other $M-m$ Pauli operators in $O$, one would need to find additional HT circuits, e.g., by making educated guesses based on careful inspection of the circuits {\crossrefcolor in Tab.~VII of SM Sec.~XIII}.
    The performance of such guesses can be easily assessed by checking for how many of the remaining Pauli operators  Eq.~\eqref{eq:linear_condition_A} is fulfilled.
    The advantage of the here-presented HT circuit over tensor product bases is quantified by $N^\text{circs}_\text{TPB}$, which is the number of circuits needed to diagonalize the same operators $P_1,\ldots,P_m$ if two-qubit gates are forbidden. See methods for details about Hamiltonians.
    }
    \begin{ruledtabular}
    \begin{tabular}{c|cccccc}
     $n$ &  20 & 40  &   60 &   80 & 100 & 120 \\ 
     $M$  % [7151, 116595, 594955, 1886543, 4611051, 9559319]
     &  $7.2\text{k}$   
     &  $117\text{k}$
     &  $595\text{k}$
     &  $1.9\text{M}$
     &  $4.6\text{M}$
     &  $9.6\text{M}$
     \\ \hline
     $m$ & 191 & 781 & 1771 & 3161 & 4951 & 7141 \\
     $N^\text{circs}_\text{TPB}$ & 19 & 24 & 25 & 28 & 30 & 32   
    \end{tabular}
    \end{ruledtabular}
    \label{tab:guessed_ht_circuits}
\end{table}

For a quantum chip whose connectivity graph $\Gamma_\text{con}$ has $n$ vertices and $e$ edges, 
there are $6^n$ and $2^e$ potential choices for $A$ and $\Gamma$, respectively.
Hence, a \solver{brute-force solver} for Eq.~\eqref{eq:linear_condition_A} would loop through all $6^n2^e$ choices, which quickly becomes infeasible due to the exponential size of the search space.
Restricting to a polynomially-large random subset gives rise to an efficient, probabilistic, \solver{restricted brute-force solver} for Eq.~\eqref{eq:linear_condition_A}, however, we empirically find that this approach has a vanishingly low success probability. 
After closely investigating the mathematics behind Eq.~\eqref{eq:linear_condition_A}, 
we can overcome this problem and formulate an efficient probabilistic solver that performs well in practice. 
Having available a huge search space will then turn into a powerful feature:
the expressivity of Eq.~\eqref{eq:linear_condition_A} enables us to construct ultra-short diagonalization circuits.

\subsection*{Mathematical Results}
\begin{figure}[t]
\centering
 
\includegraphics[width=\columnwidth]{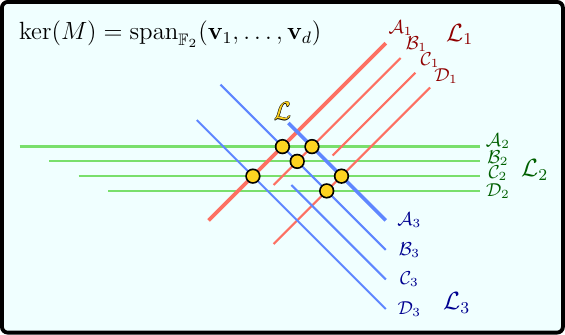}
\caption{\label{fig:geometry}  
Illustration of the solution space $\mathcal{L}$ (yellow dots) of Eq.~\eqref{eq:linear_condition_A} for a fixed graph $\Gamma$ in a hypothetical scenario with $n=3$ qubits. 
The ambient space is the $d$-dimensional null space of the binary matrix $M$ defined in Eq.~\eqref{eq:LGS}.
In this example, $\mathcal{L}$ is the intersection of  three quadric hypersurfaces $\mathcal{L}_{1}$ (red), 
$\mathcal{L}_{2}$ (green), and 
$\mathcal{L}_{3}$ (blue).
Every hypersurface $\mathcal{L}_i$ is the union of four  non-intersecting
affine subspaces $\mathcal{A}_i$, $\mathcal{B}_i$, $\mathcal{C}_i$, and $\mathcal{D}_i$ (parallel lines) defined in Eqs.~\eqref{eq:affine_subspace_a}--\eqref{eq:affine_subspace_d}.
}
\end{figure}

We have  reduced the task of constructing HT diagonalization circuits to the problem of solving Eq.~\eqref{eq:linear_condition_A} for a subgraph $\Gamma\subset \Gamma_\mathrm{con}$ and a symplectic, invertible matrix~$A$ whose blocks in Eq.~\eqref{eq:block_structure} are diagonal.
In our case, it is sufficient that $A$ is invertible because every invertible matrix $A_i\in \FF_2^{2\times 2}$, which represents $U_i$ in $U=U_1\otimes \ldots \otimes U_n$, is necessarily also symplectic, see Tab.~\ref{tab:clifford_signs}.
Because of $\det(A) = \prod_{i=1}^n \det(A_i)$  and $\FF_2=\{0,1\}$, the invertibility of $A$ is equivalent to $\det(A_1)=\ldots = \det(A_n)=1$.
%These $n$ quadratic constraints pose the biggest hurdle for solving Eq.~\eqref{eq:linear_condition_A}.
As we will show soon in Eq.~\eqref{eq:Q_definition}, for a fixed graph $\Gamma$,
it is possible to rewrite these determinants as quadratic forms
\begin{equation} \label{eq:quadric}
    \det(A_i) =  \boldsymbol{\lambda}^\mathrm{T} Q_{i} \boldsymbol{\lambda}.
\end{equation}   
%% Geometrically, % an algebraic equation of the form
The equation
$ \boldsymbol{\lambda}^\mathrm{T} Q_{i} \boldsymbol{\lambda} = 1$ defines what is known in algebraic geometry as a quadric hypersurface $\mathcal{L}_{i} \subset \FF_2^d$,
i.e., a generalization of a conic section, see Fig.~\ref{fig:geometry}.
As a consequence, $\det(A_1)=\ldots=\det(A_n)=1$  defines the intersection $\mathcal{L} = \bigcap_{i=1}^n \mathcal{L}_{i}$.
Note that the points in $\mathcal{L}$ are in one-to-one correspondence with single-qubit Clifford   layers that transform $P_1,\ldots, P_m$ into stabilizers of $\ket{\Gamma}$.
In particular, $\mathcal{L}$ is non-empty if and only if it is possible to diagonalize $P_1,\ldots,P_m$ with a circuit based on $\Gamma$ as in Fig.~\ref{fig:diagonalization_circuit}.

In our setting, every quadric hypersurface $\mathcal{L}_i$ can be written as the union of at most four affine spaces.
Therefore, their intersection  $\mathcal{L} = \bigcap_{i=1}^n \mathcal{L}_{i}$ is the union of at most $4^n$ intersections of affine subspaces, each of which is itself an affine subspace.
Using Gaussian elimination over $\FF_2$, we can efficiently probe whether such an intersection is non-empty and, once a point $\lambda \in \mathcal{L}$ is found, we have successfully constructed a HT circuit.

What is gained? 
At first glance, not much: we still have $2^e$ choices for $\Gamma \subset \Gamma_\text{con}$ and up to $4^n$ potential intersections in which a solution might lie. 
Hence, also in this reformulation, a brute force approach will be inefficient in the worst case.
It turns out, however, that the efficient but probabilistic version, where only a polynomial number of subgraphs $\Gamma \subset \Gamma_\text{con}$ and intersections is probed, has a sufficiently high empirical success probability to facilitate the construction of HT circuits for various practical problems as we will demonstrate later in Fig.~\ref{fig:all_examples}.

For pedagogical reasons, we have already revealed the geometry behind Eq.~\eqref{eq:linear_condition_A}.
Let us now delve into the corresponding algebra and show that the picture in Fig.~\ref{fig:geometry} is correct.
To solve Eq.~\eqref{eq:linear_condition_A} in practice, we have to bring $\det(A)=1$ into a form a classical computer can deal with.
First, we exploit that the blocks of $A$ in Eq.~\eqref{eq:block_structure} are diagonal.
Thus, we can replace $A$ by the vector
\begin{equation} \label{eq:vectorization_of_A}
    \mathbf{a}=\begin{bmatrix}
    \mathbf{a}^{xx}\\
    \mathbf{a}^{xz}\\
    \mathbf{a}^{zx}\\
    \mathbf{a}^{zz}
    \end{bmatrix} \in \FF_2^{4n}
\end{equation}
where the vector $\mathbf{a}^{xx}=(a^{xx}_1, \ldots, a^{xx}_n) \in \FF_2^n$ defines the $xx$-block via $A^{xx}=\diag(\mathbf{a}^{xx})$, and likewise for $\mathbf{a}^{xz}$, $\mathbf{a}^{zx}$, and $\mathbf{a}^{zz}$.
In this notation, Eq.~\eqref{eq:linear_condition_A} reads $M\mathbf{a}=0$ for the $(mn \times 4n)$-matrix 
\begin{equation}
M = 
\begin{bmatrix}
\Gamma \diag(\mathbf{r}_1) &  
\Gamma \diag(\mathbf{s}_1) &
\diag(\mathbf{r}_1) &  \diag(\mathbf{s}_1) 
\\
\vdots & \vdots & \vdots & \vdots 
\\ 
\Gamma \diag(\mathbf{r}_m) & 
\Gamma \diag(\mathbf{s}_m) & 
\diag(\mathbf{r}_m) &    \diag(\mathbf{s}_m) 
\end{bmatrix} 
.
\label{eq:LGS}
\end{equation} 
Via Gaussian elimination, we can efficiently compute a basis $\mathbf{v}_1, \ldots, \textbf{v}_d\in \FF_2^{4n}$ of the null space of $M$.
Thus, every vector $\mathbf{a}\in \FF_2^{4n}$ with $M\mathbf{a}=0$ is of the form
\begin{equation}\label{eq:ansatz_a}
\mathbf{a}=\sum_{j=1}^d \lambda_j \mathbf{v}_j
\end{equation}
for some vector $\boldsymbol{\lambda}\in \FF_2^d$.
In order to correspond to a physical solution, 
the vector $\mathbf{a}$ also needs to fulfill
\begin{equation} \label{eq:detAi}
 \det(A_i) = a^{xx}_{i} a^{zz}_{i} + a^{zx}_{i} a^{xz}_{i} = 1
\end{equation}
for every $i\in\{1,\ldots,n\}$ as this will allow us to invert the irreducible representation $\mathcal{C}_n \rightarrow \GL(\FF_2^{2n}), U\mapsto A$. 
Based on Ansatz~\eqref{eq:ansatz_a}, we find
\begin{equation}
   a^{xx}_{i} a^{zz}_{i}  
   = \sum_{j,j'=1}^d \lambda_j v_{i,j}^{xx} v_{i,j'}^{zz} \lambda_{j'} 
   = \boldsymbol{\lambda}^\mathrm{T} 
   (\mathbf{x}_i \mathbf{z}_i^\mathrm{T})
   \boldsymbol{\lambda}  
\end{equation}
and similarly $a^{xz}_{i} a^{zx}_{i}  = \boldsymbol{\lambda}^\mathrm{T}  (\mathbf{w}_i \mathbf{y}_i^\mathrm{T}) \boldsymbol{\lambda}$,
where we have introduced 
$\mathbf{x}_i=(v_{i,1}^{xx},\ldots, v_{i,d}^{xx})$,
$\mathbf{z}_i=(v_{i,1}^{zz},\ldots, v_{i,d}^{zz})$,
$\mathbf{w}_i=(v_{i,1}^{xz},\ldots, v_{i,d}^{xz})$, and 
$\mathbf{y}_i=(v_{i,1}^{zx},\ldots, v_{i,d}^{zx})$.
Further inserting these expressions into Eq.~\eqref{eq:detAi}, we obtain 
$\det(A_i) = \boldsymbol{\lambda}^\mathrm{T} (\mathbf{x}_{i} \mathbf{z}_i^{\mathrm{T}} +\mathbf{w}_{i} \mathbf{y}_i^{ \mathrm{T}}) \boldsymbol{\lambda}$ and, by defining the matrix
\begin{equation} \label{eq:Q_definition}
Q_i =  \mathbf{x}_{i} \mathbf{z}_i^{\mathrm{T}} +\mathbf{w}_{i} \mathbf{y}_i^{ \mathrm{T}}
\in \FF_2^{d \times d},
\end{equation}
we finally arrive at Eq.~\eqref{eq:quadric}.
To proceed, we point out that
$\det(A_i) =  \boldsymbol{\lambda}^\mathrm{T} Q_{i} \boldsymbol{\lambda}=1$ is equivalent to
\begin{equation} \label{eq:6cases} 
    \begin{bmatrix}
        \boldsymbol\lambda^\mathrm{T} \mathbf{x}_i \\
        \boldsymbol\lambda^\mathrm{T} \mathbf{z}_i \\
        \boldsymbol\lambda^\mathrm{T} \mathbf{w}_i \\
        \boldsymbol\lambda^\mathrm{T} \mathbf{y}_i
    \end{bmatrix}
    \in \left\{
\begin{bmatrix}  0 \\ 0 \\ 1 \\ 1   \end{bmatrix}, 
\begin{bmatrix}  0 \\ 1 \\ 1 \\ 1   \end{bmatrix}, 
\begin{bmatrix}  1 \\ 1 \\ 0 \\ 0   \end{bmatrix}, 
\begin{bmatrix}  1 \\ 1 \\ 0 \\ 1   \end{bmatrix},  
\begin{bmatrix}  1 \\ 0 \\ 1 \\ 1   \end{bmatrix},  
\begin{bmatrix}  1 \\ 1 \\ 1 \\ 0   \end{bmatrix}
\right\}.
\end{equation}
For example, if $\boldsymbol\lambda^\mathrm{T} \mathbf{x}_i=0$, the value of  $\boldsymbol\lambda^\mathrm{T} \mathbf{z}_i$ is irrelevant and we need $\boldsymbol\lambda^\mathrm{T} \mathbf{w}_i=  \boldsymbol\lambda^\mathrm{T} \mathbf{y}_i=1$ for $\det(A_i)=1$.
These are the first two cases in Eq.~\eqref{eq:6cases}.
Defining affine hyperplanes $\mathcal{X}_i^{(c)}=\{\boldsymbol\lambda \in \FF_2^d \ \vert \ \boldsymbol\lambda^\mathrm{T} \mathbf{x}_i=c\}$ for $c\in \FF_2$, and similarly $\mathcal{Z}_i^{(c)}$, $\mathcal{W}_i^{(c)}$, and $\mathcal{Y}_i^{(c)}$,
we can rephrase these two cases as $\boldsymbol{\lambda}$ being contained in 
the affine subspace
\begin{equation} \label{eq:affine_subspace_a}
\mathcal{A}_i =  \mathcal{X}_i^{(0)} \cap \mathcal{W}_i^{(1)} \cap \mathcal{Y}_i^{(1)}. 
\end{equation}
For the middle two cases in Eq.~\eqref{eq:6cases}, the roles of $(x,z)$ and $(w,y)$ are interchanged:
these cases are equivalent to $\boldsymbol{\lambda}$ being contained in the affine subspace
\begin{equation}
\mathcal{B}_i =  \mathcal{W}_i^{(0)} \cap \mathcal{X}_i^{(1)} \cap \mathcal{Z}_i^{(1)}. 
\end{equation}
Similarly, the remaining two cases in Eq.~\eqref{eq:6cases} lead to
\begin{align}
 \mathcal   C_i &= \mathcal{X}_i^{(1)} \cap \mathcal{Z}_i^{(0)} \cap \mathcal{W}_i^{(1)} \cap \mathcal{Y}_i^{(1)}  \\
  \text{and } \mathcal  D_i &= \mathcal{X}_i^{(1)} \cap \mathcal{Z}_i^{(1)} \cap \mathcal{W}_i^{(1)} \cap \mathcal{Y}_i^{(0)},
\label{eq:affine_subspace_d}
\end{align}
respectively.
Finally, by introducing
\begin{equation}  \label{eq:quadric_hypersurface_as_four_affine_spaces}
    \mathcal{L}_i = \mathcal{A}_i \cup  \mathcal{B}_i \cup  \mathcal{C}_i \cup  \mathcal{D}_i, 
\end{equation}
we obtain that $\det(A_i)=1$ is equivalent to
$\boldsymbol{\lambda} \in \mathcal{L}_i$.
We defer further technical details about how to best implement the efficient probabilistic solver based on Gaussian elimination as well as additional mathematical insights for simplifying the problem to the methods section.

\subsection*{Reformulation as an Optimization Problem}
It is possible to restate Eq.~\eqref{eq:linear_condition_A} as the feasibility problem of an 
\emph{integer quadratically constrained program} (IQP), a special case of a 
mixed integer quadratically constrained program (MIQCP)
which can be tackled with powerful numerical solvers such as Gurobi~\cite{gurobi}.
The problem instance is encoded into the binary matrices $R,S \in \FF_2^{n\times m}$, which host the exponent vectors of the Pauli operators $P_1,\ldots, P_m$ to be diagonalized, and into the connectivity graph $\Gamma_\text{con}$ with $e$ edges.
A straightforward albeit computationally expensive procedure is the following, which we refer to as \texttt{naive numerical solver}:
\begin{itemize}
    \item Introduce $4n$ free binary (Boolean) variables $\mathbf{a}=(a_1^{xx}, \ldots, a_n^{zz})$ as vectorizations of matrices $A^{xx},\ldots, A^{zz}$ as stated in Eq.~\eqref{eq:vectorization_of_A}.

    \item Introduce $e$ binary (Boolean) variables $\gamma_{i,j}$, one for each edge in the connectivity graph $\Gamma_\text{con}$.
    The final values of $\gamma_{i,j}$ determine whether or not the corresponding hardware-supported $\CZ$~gates are used.
    
    \item Introduce a matrix $N=(\nu_{j,k})
    \in \ZZ^{n \times m}$ of $n\times m$ integer slack variables and define $n m$ quadratic constraints, one for each entry in the matrix equation
    $\Gamma A^{xx}R + \Gamma A^{xz}S = A^{zx} R + A^{zz}S + 2N$.
    Note that the term $2N$ exploits Eq.~\eqref{eq:linear_condition_A} being defined over $\FF_2^{n\times m}$, i.e., equality only needs to hold modulo 2.
    
    \item Define quadratic constraints, $ a^{xx}_{i} a^{zz}_{i} + a^{zx}_{i} a^{xz}_{i} = 1$  for each $i\in\{1,\ldots, n\}$, to ensure invertibility as a determinant constraint.
    Here, there is no need to introduce additional integer slack variables that account for the fact that $\det(A_i)=1$ is supposed to hold modulo $2$ because $1$ is the only odd value that the expression $ a^{xx}_{i} a^{zz}_{i} + a^{zx}_{i} a^{xz}_{i} $ can possibly take.
    
    \item Specify a target function to be minimized, e.g., the number of two-qubit gates in the final circuit, $\text{cost}(\mathbf{a},\Gamma) = \sum_{i< j}\gamma_{i,j}$.
\end{itemize} 
In general, IQP is 
computationally hard in worst case complexity: it is in NP, while the decision version is NP-complete~\cite{pia_mixed_integer_2017}.
%MIQCP are NP-complete in worst-case complexity:
%If the decision version of MIQCP is feasible, then there exists a solution of polynomial size.
%~\cite{pia_mixed_integer_2017}.}
%
We can reduce the complexity of the naive IQP
above by leveraging our mathematical insights from the previous subsection.
Treating only one fixed subgraph $\Gamma\subset \Gamma_\text{con}$ at a time, we can compute $Q_1,\ldots,Q_n$ via Eq.~\eqref{eq:Q_definition}. 
Here, however, we regard the entries of $Q_i$ as real numbers rather than equivalence classes of integers modulo~$2$.
As a direct alternative to intersecting affine subspaces, 
we can attempt to find a solution $\boldsymbol{\lambda}\in \mathcal{L}$ by executing the following IQP which we call \texttt{informed numerical solver}:
\begin{itemize}
    \item Introduce $d$ binary variables $\boldsymbol{\lambda}=(\lambda_1,\ldots, \lambda_d)$ as well as $n$ integer slack variables $\mu_1,\ldots, \mu_n$.
    \item Define $n$ quadratic constraints, $\boldsymbol{\lambda}^\mathrm{T} Q_{i} \boldsymbol{\lambda} = 1 + 2\mu_i$.
   % \item If necessary, specify a target function to be minimized, e.g., $\text{cost}(\boldsymbol{\lambda})=0$.
    \item If necessary, specify a trivial target function. % to be minimized, e.g., $\text{cost}(\boldsymbol{\lambda})=0$.
\end{itemize}
Note that, by considering only one subgraph $\Gamma$ at a time,
the quadratic constraints in Eq.~\eqref{eq:linear_condition_A} become linear constraints over Boolean variables.
This allows us to efficiently identify the ambient space in Fig.~\ref{fig:geometry}, however, 
the constraints ensuring invertibility of $A$ still define quadratic constraints, which can be cast into the form of an IQP.
The structured problems encountered here feature only $n$ quadratic constraints and can be solved in practice for larger system sizes.
Empirically, we will demonstrate later in Fig.~\ref{fig:all_examples} that highly-optimized, commercially available MIQCP solvers
constitute a viable alternative to our provably-efficient algebraic solver.
\begin{table}[t] 
    \centering
    \begin{ruledtabular}
    \begin{tabular}{c|c|c}
        name & runtime & advantage  \\ \hline
        \solver{BF} & exp. & yields conclusive answer \\ 
        \solver{restr.~BF} & poly. & - \\ \hline
        \solver{exh.~alg.} & exp. & as BF but often faster \\ 
        \solver{restr.~alg.} & poly. & performs well in practice  \\ \hline
        \solver{naive num.}  & limited by IQP &   $\CZ$ count minimization \\
        \solver{informed num.} & limited by IQP &   faster than \solver{naive num.}
    \end{tabular}
    \end{ruledtabular}
    \caption{Overview of our solvers for Equation~\eqref{eq:linear_condition_A}. 
    The \solver{brute-force (BF) solver} loops over all $6^n2^e$ choices of $(A,\Gamma)$ and has an exponential (exp.)~runtime in $n$. 
    When restricted to a polynomial subset of choices for $(A,\Gamma)$, 
    the \solver{restricted (restr.)~BF solver} has a polynomial (poly.)~runtime but a vanishingly low success probability.
    %, rendering it utterly useless.
    The  \solver{exhaustive (exh.)~algebraic (alg.)~solver} loops over all $2^e$ subgraphs of $\Gamma_\text{con}$ and, in the worst case, through all $4^n$ intersections in Fig.~\ref{fig:geometry}; it can be fast as it terminates prematurely when either a solution $\boldsymbol{\lambda}\in \mathcal{L}$ is found or $\mathcal{L}=\diameter$ is concluded.
    The \solver{restricted algebraic solver} loops over a random selection of $s(n) = \poly(n)$ subgraphs of $\Gamma_\text{con}$ and employs a cutoff $c(n)$ as described in the methods section; it performs well in practice (see Fig.~\ref{fig:all_examples}) and has poly.~runtime by design.
    The \solver{naive numerical (num.)~solver} leverages an IQP solver with binary variables $\mathbf{a}, \Gamma$ and integer slack variables $\nu_{i,j}$ as described in the main text; it is not implemented here due to the large number of variables.
    %because of the large number of variables it requires.
    The \solver{informed num.~solver} reduces the number of variables by  leveraging our knowledge about the geometry of Eq.~\eqref{eq:linear_condition_A};
    this solver loops over a random selection of $s(n) = \poly(n)$ subgraphs, for each of which it solves an IQP problem with binary variables $\lambda_j$ and integer slack variables $\mu_i$ as described in the main text.
    For both {numerical solvers}, the runtime is limited by the leveraged %MIQCP 
    IQP solver which can be fast in practice.
    }
    \label{tab:solvers_overview}
\end{table}

\subsection*{Solver Overview}

For an overview of all solvers, %  for Eq.~\eqref{eq:linear_condition_A}, 
see Tab.~\ref{tab:solvers_overview}.
In principle, each solver (except for \solver{naive num.})~can be executed in an exhaustive mode, where all  $s(n) = 2^e$ subgraphs of $\Gamma_\text{con}$ instead of only $s(n) = \poly(n)$ are probed. 
In the exhaustive mode, every solver has an exponential runtime which limits their range of applicability. 
For small system sizes (e.g., $n\lesssim 12 $ in Fig.~\ref{fig:all_examples}), we recommend using the \solver{exhaustive, informed numerical solver} as it performs best in practice.
For larger system sizes, we recommend either the \solver{restricted, informed numerical solver},
which can be very fast, or the 
\solver{restricted algebraic solver}, whose runtime can be easily controlled, see methods section ``Runtime Analysis''. 
Hereby, the hyperparameter $s(n)$ and, if applicable, also $c(n)$, should be selected appropriately, see Fig.~\ref{fig:all_examples}, methods, and {\crossrefcolor SM Sec.~VII} for further details and examples.

\subsection*{Experimental Demonstration}
\begin{figure}[t]
    \includegraphics[width=\columnwidth]{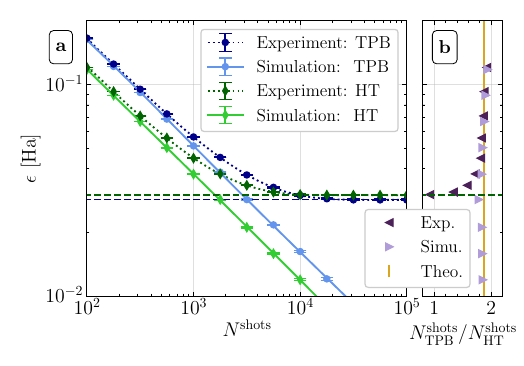} 
    \caption{Experimental shot reduction of HT readout circuits over TPBs.
    The experiment is conducted with eight superconducting qubits on \emph{ibm\_washington} and compared to an ideal Qiskit simulation~\cite{Qiskit}.
    \textbf{(a)} 
    Error $\epsilon$ (see main text) as a function of the total number of optimally-allocated shots.
    Error bars show the error on the mean of $\epsilon$ averaged over $\lfloor 5\times 10^7/N^\text{shots} \rfloor$ independent repetitions of the experiment.
    \textbf{(b)} 
    Ratio $ N^\text{shots}_\text{TPB}/N^\text{shots}_\text{HT}$ of shots needed to estimate $\langle O \rangle$ up to a total error of $\epsilon$.
    In the absence of implementation errors (Theo.), this ratio is independent of $\epsilon$ and given by $R_\text{HT}/R_\text{TPB}\approx 1.87$.
    See {\crossrefcolor SM Sec.~XIII} for further details. 
    }
    \label{fig:experiment}
\end{figure}

Next we demonstrate that our theoretical framework can enhance the efficiency of today's state-of-the-art quantum computers in practice.
A central task is to estimate the expectation value $\langle O \rangle = \Tr[\rho O]$ for an observable $O$ of interest and a state $\rho$  prepared on the quantum computer.
For a fixed shot budget (total number of available circuit executions),
the goal is to estimate $\langle O \rangle$ as accurately as possible.
As a concrete example, we consider an eight-qubit molecular Hamiltonian $O=\sum_{i=1}^M c_iP_i$ with $M=184$ Pauli operators that represents a four-atomic linear hydrogen chain, see {\crossrefcolor SM Sec.~XIII}   for details.
To enable a fair comparison of different estimation procedures, 
we restrict ourselves to the paradigm of non-overlapping Pauli groupings with reliably-executable readout circuits~\cite{wu_overlapped_grouping_2021}.
To the best of our knowledge, the previous state-of-the-art method within this paradigm is the collection of single-qubit readout circuits that arise from applying the \emph{Sorted Insertion }algorithm~\cite{crawford_efficient_quantum_2021} that groups $\{P_1,\ldots, P_{184}\}$ into, in this case, $N^\text{circs}_\text{TPB} = 35$ QWC subsets.
As an alternative, we propose to use $N^\text{circs}_\text{HT}=10$ HT readout circuits which allow us to measure the same $M=184$ Pauli operators and which contain no more than four linearly-connected $\CZ$ gates.
This reduction in the number of readout circuits is possible because being jointly-HT diagonalizable is a less stringent requirement than QWC.
What matters in the end, however, is the error $\epsilon = \vert E_\text{exp} - E_\text{ideal}\vert$ 
by which the experimentally measured energy $E_\text{exp}$ differs from the ideal one.
Hereby, one should minimize $\epsilon$ by optimally distributing the available shots among the $N_\text{TPB}^\text{circs}=35$ or $N_\text{HT}^\text{circs}=10$ readout circuits~\cite{crawford_efficient_quantum_2021}, see 
{\crossrefcolor SM Sec.~IV}.
To enable a pronounced comparison between the two readout methods, 
we select a product target state vector $\ket{\Psi}=\ket{\psi_1}\otimes \ldots \otimes \ket{\psi_8}$
that can be prepared with a high fidelity.
This state has a large energy $E_\text{ideal} = \bra{\Psi}O\ket{\Psi} = -0.49264\unitspace\text{Ha}$ compared to the ground state energy, $\min_\Phi \bra{\Phi}O\ket{\Phi}=-2.26752\unitspace\text{Ha}$,
where all reported energies include Coulomb repulsion between the nuclei.
We perform experiments for both readout methods and present the results in Fig.~\ref{fig:experiment}.
The error $\epsilon$  is plotted in Fig.~\ref{fig:experiment}\,\textbf{a} as a function of the shot budget $N^\text{shots}$ for both TPB (blue circles) and HT (green diamonds). 
We compare the experimental results (dark) to classical, noise-free simulations (bright).
In Fig.~\ref{fig:experiment}\,\textbf{b} we display the shot reduction ratio 
$ N^\text{shots}_\text{TPB}/N^\text{shots}_\text{HT}$ for achieving a target error $\epsilon$ that our HT circuits offer over conventional TPBs.
For very low budgets of a few hundred shots in total, we see that the experimental data perfectly agree with the simulation.
This is because sampling errors dominate here.
While the simulated errors generally decrease as $1/\sqrt{N^\text{shots}}$, 
the experimental errors eventually saturate at a finite bias $b$ stemming 
mainly from noise in the diagonalization circuits and the readout-error-mitigated~\cite{bravyi_mitigating_measurement_2021} $Z$-measurements. 
Unsurprisingly, $b_\text{HT} = 0.0299(2)\unitspace\text{Ha}$ is larger than  $b_\text{TPB} \approx 0.0284(2)\unitspace\text{Ha}$.
For shot budgets below $ 10^4$, however, we observe that $\epsilon$ is smaller for HT than for TPB.
Hence, our HT readout circuits outperform TPBs in the low-shot regime, which is of practical importance, see {\crossrefcolor SM Sec.~I}.

\section*{Discussion}
In the remainder of this article, we discuss the reusability potential of HT Pauli groupings, investigate the performance of our solvers, and highlight further potential use cases for our theoretical framework.

%\subsection*{Reusability of Pauli Groupings}
 
\begin{figure}[t] 
    \includegraphics[width=\columnwidth]{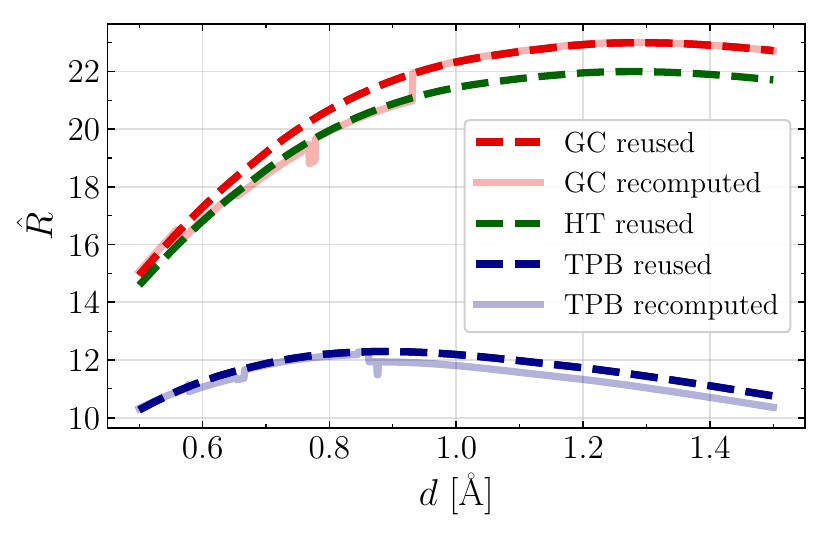} 
    \caption{Reusability opportunity. Estimated shot reduction $\hat{R}$ defined in Ref.~\cite{crawford_efficient_quantum_2021} for measuring the energy $\langle O \rangle$ of a hydrogen chain Hamiltonian as a function of the interatomic distance $d$ between the hydrogen atoms.
    Due to continuity, there is no need to regroup the Pauli operators in $O$ when $d$ is updated.
    This enables enormous savings in preprocessing cost.
    }
    \label{fig:reusability}
\end{figure}

\begin{figure*}[t]
    \includegraphics{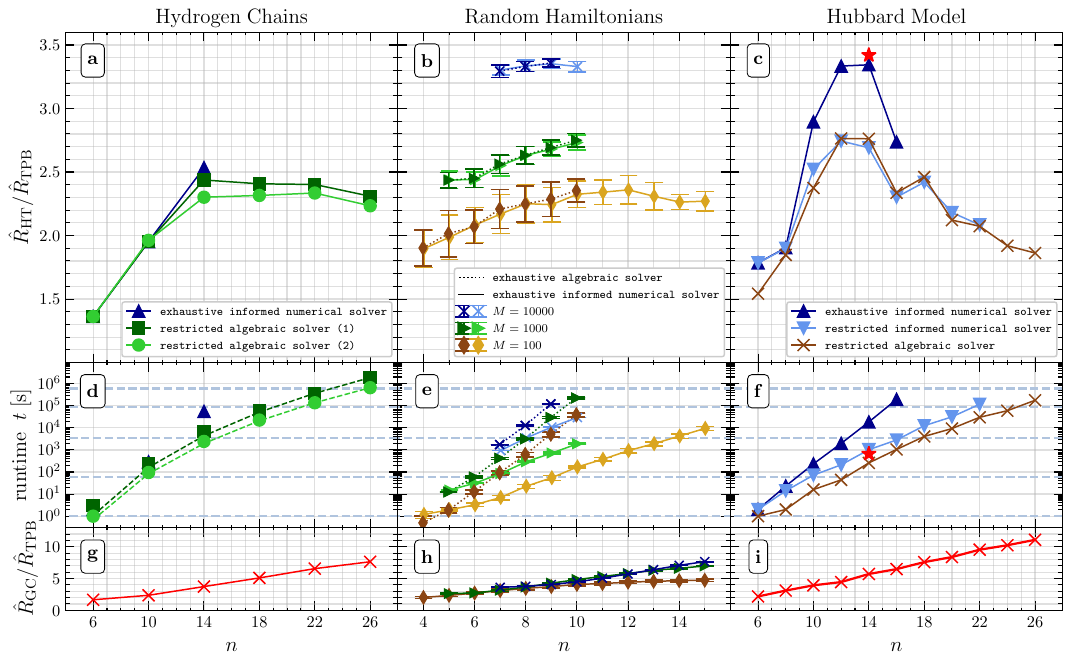}
    \caption{Performance of a greedy Pauli grouping algorithm with various solvers from Tab.~\ref{tab:solvers_overview}. 
    \textbf{(a-c)} Estimated shot reduction $\hat{R}_\text{HT}/\hat{R}_\text{TPB}$ for three classes of $n$-qubit Hamiltonians $O$.
    The HT readout circuits assume a linear hardware connectivity.
    Different curves correspond to different solvers from Tab.~\ref{tab:solvers_overview} and, in the case of random Hamiltonians (\textbf{b,e,h}), to
    different numbers $M$ of Pauli operators in $O$.
    For the single data point with the red star symbol ({\color{red}$\bigstar$}) at $n=14$ in panels \textbf{c} and \textbf{f}, we use the \solver{restr.~alg.~solver} with a fine-tuned choice of hyperparameters. 
   \textbf{(d-f)}  
    Runtime of our HT Pauli grouping algorithm.
    {\textbf{(g-i)} Estimated shot reduction $\hat{R}_\text{GC}/\hat{R}_\text{TPB}$  that is in principle achievable with unrestricted Clifford circuits if two-qubit errors are neglected, which is an unrealistic assumption for near-term quantum devices.}
    Therefore, HT readout circuits present the best viable option in this comparison.
    See methods for further details.
    }
    \label{fig:all_examples}
\end{figure*}

Every grouping of a fixed set $\{P_1, \ldots, P_M\}$ into jointly-diagonalizable subsets has vast reusability potential:
for a fixed observable $O=\sum_{i=1}^M c_i P_i$, one can estimate $\Tr[\rho O]$ for countless experimental states $\rho$,
e.g., different eigenstates of $O$ as well as the myriad of VQE trial states that are encountered until such eigenstates are found.
Similarly, in simulations of chemical reactions, $\Tr[\rho(t)O]$ is mapped out for a time series of quantum states $\rho(t)$.

Moreover, for any linear combination $O'=\sum_{i=1}^M c'_i P_i$ of the same Pauli operators, it is possible to estimate $\langle O'\rangle$ with the same readout circuits as for $\langle O \rangle$.
Therefore, a single Pauli grouping can suffice to map out low-energy Born-Oppenheimer surfaces via, e.g., VQE.
At the same time, we are guaranteed that the quality of the reused grouping does not deteriorate abruptly as the estimated shot reduction $\hat{R}$
depends continuously on the nuclear coordinates~\cite{crawford_efficient_quantum_2021}.

We illustrate this fact in Fig.~\ref{fig:reusability} for the example of an eight-qubit Hamiltonian $O(d) =\sum_{i=1}^{184} c_i(d)P_i$ that represents a four-atomic linear hydrogen chain for which the interatomic spacing $d$ is varied.
For different Pauli groupings of $O(d)$, we plot $\hat{R}$, which should be regarded as a state-independent estimate of the shot reduction ratio $N^\text{shots}_\text{GPM}/N^\text{shots}_\text{IPM}$ (cf.~Fig.~\ref{fig:experiment}\,\textbf{b}),
where $N^\text{shots}_\text{GPM}$ and $N^\text{shots}_\text{IPM}$, respectively, denote the number of shots required to measure $\langle O(d)\rangle$ to a fixed precision $\epsilon$ if grouped Pauli measurements (GPM) and individual Pauli measurements (IPM) are performed.
For example, $O(d=1.0\unitspace\text{\AA})$ coincides with the Hamiltonian from Fig.~\ref{fig:experiment} and
the estimated shot reduction ratio of HT over TPB, $\hat{R}_\text{HT}/\hat{R}_\text{TPB} \approx 1.76$, is close to the state-dependent value, ${R}_\text{HT}/{R}_\text{TPB} \approx 1.87$.
As explained above, we can see in Fig.~\ref{fig:reusability} that $\hat{R}$ depends continuously on $d$ if a fixed Pauli grouping is reused (dark dashed lines).
Also shown is the piecewise continuous dependence of $\hat{R}_\text{GC}$ and $\hat{R}_\text{TPB}$ if a new Pauli grouping is recomputed for every value of $d$ (bright solid lines), where the jumps unveil the values of $d$ at which the output of the Sorted Insertion algorithm changes.
Since Sorted Insertion is a greedy algorithm, recomputing the Pauli grouping can even diminish the value of $\hat{R}$.

Finally, note that Pauli groupings can be reused for numerous other applications in the context of parameter-shift rules~\cite{schuld_evaluating_analytic_2019, sweke_stochastic_gradient_2020, meyer_a_variational_2021, hubregtsen_single_component_2022}.
%\subsection*{Performance of the Proposed Solvers}
It is straightforward to adapt Sorted Insertion to HT Pauli groupings, 
see {\crossrefcolor SM Sec.~V}.
Instead of checking if a set of Pauli operators (qubitwise) commutes, one has to construct a HT readout circuit using one of the solvers from Tab.~\ref{tab:solvers_overview}.
In Fig.~\ref{fig:all_examples}, we investigate the performance of our various solvers by applying a modified Sorted Insertion algorithm to three classes of paradigmatic Hamiltonians.
In all cases, we observe in Fig.~\ref{fig:all_examples}\,\textbf{a-c} that the estimated shot reduction ratio, $\hat{R}_\text{HT}/\hat{R}_\text{TPB}$,
takes values in between $1.3$ and $3.5$.
Therefore, our HT readout circuits consistently outperform conventional TPBs.

While Sorted Insertion has a runtime of $\mathcal{O}(M^2n)$, the runtime of our adaptation is given by $\mathcal{O}(M^2 f(n))$, where $M$ still denotes the number of Pauli operators in the observable $O$, and $f(n)$ is the complexity of the selected solver,
e.g., $f(n)= \mathcal{O}(2^e6^n n^3)$ for the \solver{brute-force algebraic solver} and $f(n)=\mathcal{O}(\text{poly}(n))$ for the \solver{restricted algebraic solver}, {see methods section ``Runtime Analysis''.}
For molecular hydrogen chain Hamiltonians  for example,
two polynomial fits (green dashed lines) in Fig.~\ref{fig:all_examples}\,\textbf{d} reveal total runtimes 
$t^{(1)} \propto n^{9.8}$ and 
$t^{(2)} \propto n^{9.4}$.
Since the number of Pauli operators is here given by $M\approx 0.17 \times n^{3.68}$,
we can conclude that the \solver{restricted algebraic solver} has an empirical runtime $f(n)$ lying somewhere in between $\mathcal{O}(n^{2.0})$ and $\mathcal{O}(n^{2.4})$, which is in agreement with the expected runtime of $f(n)=\mathcal{O}(n^3)$,
see methods.

For the example of random Hamiltonians, we compare in Fig.~\ref{fig:all_examples}\,\textbf{b} the \solver{algebraic solver} with  the \solver{informed numerical solver} from Tab.~\ref{tab:solvers_overview}, both in combination with an exhaustive (exh.)~search over all $2^e$ subgraphs $\Gamma \subset \Gamma_\text{con}$.
We find that both solvers produce HT Pauli groupings of the same quality, which we believe to be near optimal for the assumed linear connectivity constraint.
By construction, both exhaustive solvers are inefficient, however, we observe in Fig.~\ref{fig:all_examples}\,\textbf{e} that the \solver{numerical solver} is much faster in practice.
This is unsurprising as this option leverages a highly-optimized MIQCP solver.

Since the \solver{exhaustive numerical solver} is our best option for constructing near-optimal HT Pauli groupings, 
we can now turn to the question about how much the quality (as measured by $\hat{R}_\text{HT}$) of the HT Pauli grouping decreases when we replace the \solver{brute-force solver} by our provably-efficient \solver{restricted algebraic solver}.
For the example of the Hubbard model 
we observe in Fig.~\ref{fig:all_examples}\,\textbf{c} that both the \solver{restricted numerical solver} and the \solver{restricted algebraic solver} produce HT Pauli groupings with the same qualitative behavior of $\hat{R}$ as the (near-optimal) \solver{exhaustive numerical solver}.
For all solvers, $\hat{R}$ first increases with the number of qubits $n$, before it drops again.
We attribute this saturation effect to the linear hardware connectivity constraint because $\hat{R}_\text{GC}/\hat{R}_\text{TPB}$ continues to grow as shown in  Fig.~\ref{fig:all_examples}\,\textbf{i} (also see Fig.~1 in {\crossrefcolor SM Sec.~II}).
Whether a similar growth is retained with 2-dimensional HT readout circuits deserves further investigation.
For $n=14$, the \texttt{exhaustive numerical solver} constructs a grouping with $\hat{R}_\text{HT}/\hat{R}_\text{TPB} \approx 3.34$ %  $\hat{R}_\text{HT}\approx 12.3$ 
after 5~hours. 
By carefully balancing the hyperparameters of the \solver{restricted algebraic solver} (see methods), we find an even better grouping with $\hat{R}_\text{HT}/\hat{R}_\text{TPB} \approx 3.42$ %  $\hat{R}_\text{HT}\approx 12.6$
in only 11~minutes (red star).

The discussion above shows that the classical bottleneck of our HT Pauli grouper can be overcome.
To corroborate this claim,
we propose and test a rudimentary HT Pauli grouper (Algorithm~2 in {\crossrefcolor SM Sec.~VI})
for which the enabled runtime savings outweigh the classical preprocessing cost for a 52-qubit example.
This demonstrates that the advantage of HT readout circuits can be scaled up to meaningful system sizes.
Finally, let us point out that our approach also works for more sophisticated molecular basis sets, see SM Sec.~XIV.

%\subsection*{Remarks}

In this article, we have introduced a theoretical framework for the construction of diagonalization circuits that can be tailored to any given hardware connectivity.
Our starting point was the observation of the fact that every set of commuting Pauli operators can be cast into the stabilizer group of a stabilizer state which is local-Clifford equivalent to a graph state.
The Pauli operators can thus be diagonalized with a quantum circuit that completely  avoids $\textsc{Swap}$ gates whenever this graph state matches the connectivity of the quantum computer.
We derived an algebraic criterion for the existence of such \emph{hardware-tailored} (HT) diagonalization circuits and introduced solvers for their construction.
An important empirical observation is that in many cases it is not necessary to apply unconstrained diagonalization circuits because also HT circuits can be constructed.
For super- and semiconducting chips, for which the connectivity graph has a bounded degree,
HT circuits have a constant depth.
In comparison to state preparation circuits such as
UCCSD, which have a gate count of $\mathcal{O}(n^4)$ and a gate depth of $\mathcal{O}(n^3)$,
the circuit complexity of HT readout circuits is therefore negligible~\cite{mcardle_quantum_computational_2020}.
The construction of HT circuits can be computationally demanding but is worthwhile given their reusability potential.
Finally, we have demonstrated the advantage of our approach over previous ones both in theory and in experiment.

There are multiple ways to further improve the efficiency of our  solvers for the construction of HT diagonalization circuits, see methods section ``Ideas for Improving our Solvers''.
Furthermore, it would be worthwhile to combine the paradigm of HT readout circuits with complementary state-of-the-art methods for lowering shot requirements such as 
iterative measurement allocation and iterative coefficient splitting~\cite{yen_deterministic_improvements_2023}. 

Lowering shot requirements, e.g., in variational algorithms as discussed above, is just one of many potential applications for HT diagonalization circuits.
Other, equally important use cases arise in the contexts of classical shadows~\cite{huang_predicting_many_2020}, Hamiltonian time simulation~\cite{vandenberg_circuit_optimization_2020,faehrmann_randomizing_multi_2022, anand_leveraging_commuting_2023}, and quantum error correction~\cite{lidar_quantum_error_2013}.
For instance, an encoding circuit for a stabilizer quantum error-correcting code is the same as a time-reversed diagonalization circuit for its stabilizer group.
For an outline how Hamiltonian exponentiation could potentially benefit from our HT diagonalization circuits, see {\crossrefcolor SM Sec.~IX}.

\section*{Methods}

\subsection*{ Choices of Hamiltonians}

All molecular hydrogen chain Hamiltonians were computed in the STO-3G minimal basis using Qiskit nature~\cite{Qiskit}, either in combination with pyquante~\cite{muller_pyquante_python}
or PySCF~\cite{sun_pyscf_the_2018}.
More sophisticated basis sets are discussed in SM Sec.~XIV.
For Tab.~\ref{tab:guessed_ht_circuits}, Fig.~\ref{fig:experiment}, and Fig.~\ref{fig:reusability},
we use pyquante and the Bravyi-Kitaev  fermion-to-qubit mapper to obtain $n$-qubit Hamiltonians representing a hydrogen chain with $\tfrac{n}{2}$ nuclei at an equidistant spacing of $d$, where we fix $d=1.0\unitspace\text{\AA}$ in Tab.~\ref{tab:guessed_ht_circuits} and Fig.~\ref{fig:experiment}.
The exact numbers of Pauli operators $M(n)$ of the $n$-qubit Hamiltonians in Tab.~\ref{tab:guessed_ht_circuits} are given by
$M(20)  =     7,150$, 
$M(40)  =   116,594$,
$M(60)  =   594,954$, 
$M(80)  = 1,886,542$, 
$M(100) = 4,611,050$, and
$M(120) = 9,559,318$.
For the hydrogen chain Hamiltonians in the left column of Fig.~\ref{fig:all_examples}, on the other hand,
we use PySCF and the parity-encoding fermion-to-qubit mapper, resulting in Hamiltonians with only $n=2n_\text{at}-2$ qubits, where $n_\text{at}$ is the number of hydrogen atoms; also here the interatomic spacing is chosen as $d=1.0\unitspace\text{\AA}$.

In Fig.~\ref{fig:all_examples}\,\textbf{b}, we average over twenty random Hamiltonians of the form  $O=\sum_{i=1}^Mc_iP_i$ for each choice of qubit number $n$ and number of Pauli operators $M$.
For every $O$, the  Pauli operators $P_i \in \{I,X,Y,Z\}^{\otimes n} \backslash \{I^{\otimes n}\}$ and coefficients $c_i\in[-1,1]$ are drawn uniformly at random.
Error bars show one standard deviation.
For an in-depth investigation of the results, see {\crossrefcolor SM Sec.~VIII}.

Figure~\ref{fig:all_examples}\,\textbf{c} features the Hubbard model of $L$ fermionic modes $\hat c_{{k_j},\sigma}^{\dagger}$ with momentum $k_j = 2\pi j /L$  and spin $\sigma$ on a 1-dimensional  lattice with $L$ sites and periodic boundary conditions.
Its Hamiltonian is given by
\begin{align}
    O =&  \frac{U}{L} \sum_{i,j,l=1}^L \hat c_{k_i-k_l, \uparrow}^\dagger \hat c_{k_j+k_l, \downarrow}^\dagger  \hat c_{k_j, \downarrow}  \hat c_{k_i, \uparrow} \\
    &+\sum_{j=1}^L \sum_{\sigma\in \{\uparrow, \downarrow\}} 
    \epsilon_{k_j}   \hat c_{{k_j},\sigma}^{\dagger}  \hat c_{{k_j},\sigma}
    \nonumber
\end{align}
where $U\ge 0$ is the Coulomb energy, $\epsilon_{k_j}= -2t \cos(k_j)$ is the dispersion relation for non-interacting ($U=0$) fermions, and $t\ge0$ is the hopping strength~\cite{essler_the_one_2005}. 
We use the block-spin Jordan-Wigner encoding to convert this Hamiltonian into a linear combination of Pauli operators.
In Fig.~\ref{fig:all_examples}, we assume $U=1$ and $t=1$ to obtain concrete values for $\hat{R}$.

\subsection*{Choices of Hyperparameters}
Here we report the choices of hyperparameters of our solvers that were used to create Fig.~\ref{fig:all_examples}.
In all cases, we tailor the readout circuits to a linear connectivity by applying Algorithm~1 from {\crossrefcolor SM Sec.~V}.
The total runtime and the quality of the resulting Pauli grouping  depends on the choice of the solver from Tab.~\ref{tab:solvers_overview} and its hyperparameters.
Unless specified otherwise, all computations were carried out on an 18 core Intel Xeon CPU E5-2697 v4 @2.30GHz device~\footnote{Intel and Intel Xeon are trademarks or registered trademarks of Intel Corporation or its subsidiaries in the United States and other countries.}.

The \solver{exhaustive informed numerical solver} (dark blue triangles in Fig.~\ref{fig:all_examples}\,\textbf{d},\textbf{f} and dark markers with dotted lines in Fig.~\ref{fig:all_examples}\,\textbf{e}) searches over all $2^{n-1}$ circuit templates $\Gamma \subset \Gamma_\text{con}$ and constructs readout circuits (one for each subgraph $\Gamma$) by solving the informed IQP problem (if possible).
Then, the best circuit is chosen to assign a collection of Pauli operators to a joint measurement, see {\crossrefcolor SM Sec.~V} for further details.
The \solver{exhaustive algebraic solver} (bright markers with solid lines in Fig.~\ref{fig:all_examples}\,\textbf{e}) similarly searches over all $2^{n-1}$ circuit templates but applies the \solver{algebraic brute-force solver} instead of the \solver{informed numerical solver}.
Both exhaustive solvers have an exponential runtime because they take all subgraphs $\Gamma \subset \Gamma_\text{con}$ into account.

The \solver{restricted numerical solver} has one hyperparameter: the number $s(n)$ of subgraphs $\Gamma \subset \Gamma_\text{con}$ that are probed. 
In our current implementation, the specific choice of subgraphs is taken uniformly at random.
For the Hubbard model (Fig.~\ref{fig:all_examples}\,\textbf{f}), we work with $s(n)=\min\{1.5 n^2 - 0.5{n}, 2^{n-1}\}$ random subgraphs.

The \solver{restricted algebraic solver} has two hyperparameters: the number $s(n)$ of random subgraphs and the cutoff $c(n)$ defined before Eq.~\eqref{eq:cutoff}.
For the Hubbard model (Fig.~\ref{fig:all_examples}\,\textbf{f}), we work with $s(n)=\min\{ n^2, 2^{n-1}\}$ and $c(n)=\lfloor \log_2(n)\rfloor$,
except for the fine-tuned parameter choice (red star) where the number of subgraphs is increased from $s(n) = n^2=196$ to $s=1000$.
For hydrogen chains (Fig.~\ref{fig:all_examples}\,\textbf{d}), on the other hand, we use two different choices of constant hyperparameters:
\begin{enumerate}[(1)]
    \item $s(n)= 2000$ and $c(n)=5$.
    \item $s(n)= 1000$ and $c(n)=3$.
\end{enumerate}

Besides HT Pauli groupings, we also compute GC Pauli groupings in Figs.~\ref{fig:reusability} and~\ref{fig:all_examples} and groupings of $O$ into TPBs in Figs.~\ref{fig:experiment}--\ref{fig:all_examples}.
These groupings were obtained by applying the Sorted Insertion algorithm~\cite{crawford_efficient_quantum_2021} with the insertion conditions ``commute'' and ``qubitwise commute'', respectively.

For additional numerical investigations about how the choice of hyperparameters influences the performance of the HT Pauli grouping algorithm, see {\crossrefcolor SM Secs.~VII} and~XIV.

\subsection*{Technical Details on the Algebraic Solver}

Here we continue our technical discussion about how to algebraically construct HT diagonalization circuits.
For every qubit $i\in\{1,\ldots,n \}$,
there are three cases how the hypersurface $\mathcal{L}_i= \{\boldsymbol{\lambda} \in \FF_2^d \ \vert \ \boldsymbol{\lambda}^\mathrm{T} Q_i \boldsymbol{\lambda}  =1 \}$ could look like:
since the image of the $\FF_2$-linear map defined by the matrix $Q_i$ is contained in the span of $\mathbf{x}_i$ and $\mathbf{w}_i$,
the rank of $Q_i$ can only take one of the three values: $0$, $1$, and $2$.
If $\rank_{\FF_2}(Q_i)=0$, i.e., $Q_i=0$, then $\mathcal{L}_i = \diameter$ is empty, which implies $\mathcal{L} = \bigcap_{i=1}^n \mathcal{L}_i = \diameter$ and proves the non-existence of a HT diagonalization circuit for the fixed choice of $P_1,\ldots,P_m$ and $\Gamma$.
On the other hand, if $\rank_{\FF_2}(Q_i)=1$, e.g., $\mathbf{x}_i \mathbf{z}_i^\mathrm{T} = 0$ but $\mathbf{w}_i \mathbf{y}_i^\mathrm{T} \neq 0$,
then we need $\boldsymbol{\lambda}^\mathrm{T} \mathbf{w}_i \mathbf{y}_i^\mathrm{T} \boldsymbol{\lambda}=1$, which is equivalent to $\boldsymbol\lambda^\mathrm{T} \mathbf{w}_i= \boldsymbol\lambda^\mathrm{T} \mathbf{y}_i=1$. 
Thus, this case is degenerate and the hypersurface collapses to an affine subspace
\begin{equation}\label{eq:affine_subspace_intersection}
    \mathcal{L}_i = \mathcal{W}_i^{(1)} \cap \mathcal{Y}_i^{(1)}.
\end{equation}
For a detailed explanation about how to compute intersections of affine subspaces numerically, see {\crossrefcolor SM Sec.~X}. 
Finally, in the most general case of $\rank_{\FF_2}(Q_i)=2$, the hypersurface $\mathcal{L}_i$ defies a description simpler than the union of four affine subspaces as in Eq.~\eqref{eq:quadric_hypersurface_as_four_affine_spaces}.
As we show next, the occurrence of rank-$1$ hypersurfaces can greatly simplify the problem.

Denote the number of rank-$2$ hypersurfaces by $k$.
In our search for a point $\boldsymbol{\lambda}\in \mathcal{L}$, we first compute the intersection of all rank-$1$ hypersurfaces $\mathcal{L}_i$. 
After potentially relabelling some of the qubits, we can assume  
$Q_i= \mathbf{x}_{i} \mathbf{z}_i^{\mathrm{T}}$
and 
$Q_j = \mathbf{w}_{j} \mathbf{y}_j^{ \mathrm{T}}$ 
for all $i\in \{1,\ldots, l\}$ and $j\in \{l+1, \ldots,  n-k\}$.
Then, the intersection of all rank-1 quadric hypersurfaces is given by
\begin{align} \nonumber
\bigcap_{i=1}^{n-k} \mathcal{L}_i 
&= 
\left \{ 
\boldsymbol{\lambda}\in \FF_2^d
\
\bigg \vert 
\ 
\begin{minipage}{11.5em} 
\scalebox{.75}{$\boldsymbol{\lambda}^\text{T} \mathbf{x}_i = \boldsymbol{\lambda}^\text{T} \mathbf{z}_i =   \boldsymbol{\lambda}^\text{T} \mathbf{y}_j = \boldsymbol{\lambda}^\text{T} \mathbf{w}_j = 1 $} \\ 
\scalebox{.75}{for all $1\le i \le l <  j \le n-k$}
\end{minipage}
\right\}
 \\
&=
\left \{ 
\boldsymbol{\lambda}\in \FF_2^d 
\  \hspace{0.275mm}
\Big \vert 
\ 
C\boldsymbol{\lambda} = \mathbf{1}
\right\},
\label{eq:affine_subspace_intersection_C}
\end{align}
where $\mathbf{1}=[1,\ldots,1]^\text{T}$ and for the ($2(n-k)\times d $)-matrix
$C = [\mathbf{x}_1, \mathbf{z}_1,\ldots ,\mathbf{x}_l, \mathbf{z}_l, \  \, \mathbf{y}_{l+1},\mathbf{w}_{l+1},\ldots,\mathbf{y}_{n-k},\mathbf{w}_{n-k}]^\text{T}$
 whose rows are given by the row vectors $\mathbf{x}_1^\text{T}$ etc.
Using Gaussian elimination over $\FF_2$, we can compute the reduced row-echelon form (RREF) of the extended matrix $[C, \mathbf{1}] \in \FF_2^{2(n-k)\times(l+1)}$. 
If the last column of the RREF is a pivot column, we are in the solutionless case $\mathcal{L}=\diameter$. 
Otherwise, this column is an offset vector for $\mathcal{L}_1\cap\ldots \cap \mathcal{L}_{n-k}$, while every non-pivot column of the RREF of $C$ yields a basis vector in the standard way of linear algebra, 
see {\crossrefcolor SM Sec.~X} for technical details.

Finally, we turn to the computationally most demanding part that deals with the rank-2 hypersurfaces $\mathcal{L}_{n-k+1}, \ldots, \mathcal{L}_n$.
Similar to the matrix $C$ in Eq.~\eqref{eq:affine_subspace_intersection_C}, we introduce a  ($4k\times d$)-matrix
%\begin{equation}
$   C' = [\mathbf{x}_{n-k+1}, \mathbf{z}_{n-k+1}, \mathbf{w}_{n-k+1}, \mathbf{y}_{n-k+1},  \ldots, \mathbf{x}_{n},\mathbf{z}_{n},\mathbf{w}_{n},\mathbf{y}_{n}]^\text{T}$,
%\end{equation}
i.e., for every $j\in\{1,\ldots,k\}$, the four rows  of $C'$  from row $4j-3$ to row $4j$ are given by
$\mathbf{x}_{n-k+j}  ^\mathrm{T}$,
$\mathbf{z}_{n-k+j}  ^\mathrm{T}$,
$\mathbf{w}_{n-k+j}  ^\mathrm{T}$, and
$\mathbf{y}_{n-k+j}  ^\mathrm{T}$.
Then, a vector $\boldsymbol{\lambda}\in \FF_2^d$ is contained in $\mathcal{L}_{n-k+1} \cap \ldots \cap \mathcal{L}_n$
if and only if 
    $C' \boldsymbol{\lambda} = \mathbf{b}_{i_1}\oplus \ldots \oplus \mathbf{b}_{i_k}$
for some $i_1,\ldots, i_k \in \{1,\ldots, 6\}$, where the vectors $\mathbf{b}_1, \ldots, \mathbf{b}_6$ in the direct sum
\begin{equation}
 \mathbf{b}_{i_1}\oplus \ldots \oplus \mathbf{b}_{i_k} =
    \begin{bmatrix}
    \mathbf{b}_{i_1} \\
    \vdots \\
    \mathbf{b}_{i_k} \\
    \end{bmatrix}
\end{equation}
are the six vectors from Eq.~\eqref{eq:6cases}, i.e., 
$\mathbf{b}_1 = [0,0,1,1]^\mathrm{T}$ etc.
If it is our goal to unambiguously ascertain whether or not $\mathcal{L}$ is empty, we have to check an exponential number of cases.
This is only feasible for small qubit numbers $n$ or for graphs with small components, as we explain in {\crossrefcolor SM Sec.~XII}.
To save computing time, we treat all combinations of the vectors $\mathbf{b}_i$ at once by introducing ($4\times 6^j$)-matrices
\begin{equation}
    B_j  = [\underbrace{\mathbf{b}_1,\ldots, \mathbf{b}_1}_{6^{j-1} \text{ times}}, \ldots , \underbrace{\mathbf{b}_6,\ldots, \mathbf{b}_6}_{6^{j-1}  \text{ times}}] 
\end{equation}
for $j\in\{1,\ldots, k\}$ and using them as blocks for enlarging $C'$ to a matrix $B \in \FF_2^{4k \times (d + 6^k)}$.
Hereby, the four rows from row $(4j-3)$ to row $4j$ of $B$ are given by
\begin{equation}
\left[
\begin{minipage}{4em} \centering
$\mathbf{x}_{n-k+j}  ^\mathrm{T}${\color{white},}  \\
$\mathbf{z}_{n-k+j}  ^\mathrm{T}${\color{white},}  \\
$\mathbf{w}_{n-k+j}  ^\mathrm{T}${\color{white},} \\
$\mathbf{y}_{n-k+j}  ^\mathrm{T}$,
\end{minipage}
\ 
\begin{minipage}{8em}
\centering
\Large
$\underbrace{B_j, \ldots , B_j} _{6^{k-j} \text{\normalsize{} times}}$
\end{minipage}
\right]
\in \FF_2^{4 \times (d + 6^k) }.
\label{eq:big_lgs_4rows}
\end{equation}
In other words, $B$ is the matrix that arises from $C'$ by appending all vectors of the form $\mathbf{b}_{i_1}\oplus \ldots \oplus \mathbf{b}_{i_k}$ as additional columns.
Next, we use Gaussian elimination to bring $B$ to a row-echelon form; here, time can be saved as it is not necessary to compute the RREF of $B$.
This reveals the non-pivot columns of $B$.
Every  non-pivot column of the form $\mathbf{b}_{i_1} \oplus \ldots \oplus \mathbf{b}_{i_k}$
indicates the existence of at least one vector $\boldsymbol{\lambda}\in \mathcal{L}_{n-k-1}\cap \ldots \cap \mathcal{L}_n$.
However, we are looking for a $\boldsymbol{\lambda}$
that also lies in the affine space $\mathcal{L}_1\cap\ldots\cap \mathcal{L}_{n-k}$.
To accomplish this, we start by computing a basis and an offset vector of the entire affine space 
\begin{equation}
\left\{ \boldsymbol{\lambda}\in \FF_2^d \ \vert \ C'\boldsymbol{\lambda} =   \mathbf{b}_{i_1} \oplus \ldots \oplus \mathbf{b}_{i_k}
\right\}.
\label{eq:affine_space_cases}
\end{equation}
Then, we use the procedure explained {\crossrefcolor in SM Sec.~X} to compute the intersection of the two affine subspaces in Eqs.~\eqref{eq:affine_subspace_intersection_C} and~\eqref{eq:affine_space_cases}.
Since this results in a subset of $\mathcal{L}$, we can finish if we find a non-pivot column ${\mathbf{b}_{i_1}\oplus\ldots\oplus \mathbf{b}_{i_k}}$ in the right part of $B$ for which this intersection is not empty.
Otherwise, if this approach fails for all non-pivot columns, 
we can finally infer $\mathcal{L}=\diameter$.
In any case, we obtain a conclusive answer whether or not a layer of single-qubit Clifford gates exists
such that the corresponding graph-based circuit (see main text, Fig.~\ref{fig:diagonalization_circuit}) diagonalizes the given set of commuting Pauli operators.

\subsection*{Restricting the Algebraic Solver}

In the brute force approach of the previous subsection to algebraically solve Eq.~\eqref{eq:linear_condition_A} from the main text, we iterate through a number of affine subspaces that grows exponentially in the number $k\le n$ of qubits $i$ for which $\mathcal{L}_i$ is a rank-2 quadric hypersurface.
For large problem sizes, this is infeasible and we can instead restrict the search to a smaller number $c(n)\le k$ of rank-2 quadric hypersurfaces.
For example, we can work with a constant cutoff
\begin{equation} \label{eq:cutoff}
    c(n) = \text{const.}
\end{equation}
or with a logarithmically-growing cutoff
\begin{equation} \label{eq:cutoff_log}
    c(n) = \text{const.} \times  \lfloor \log (n) \rfloor.
\end{equation}
This restriction turns our \solver{algebraic solver} (for attempting to find a solution $\boldsymbol{\lambda} \in \mathcal{L}$) into an efficient but probabilistic algorithm because only a polynomial number of intersections of affine subspaces will be probed,
see methods section ``Runtime Analysis'' below.
If no solution is found this way, we treat this case as if $\mathcal{L}$ was empty, i.e., we skip the current subgraph in our Pauli grouping algorithm from {\crossrefcolor SM Sec.~V}.

We can incorporate the cutoff $c= c(n)$ by replacing the matrix $B\in\FF_2^{4k \times (d+ 6^k)}$ in Eq.~\eqref{eq:big_lgs_4rows}
by a smaller matrix $B' \in \FF_2^{(4c+3(k-c)) \times (d+ 6^c)} $.
The first $4c$ rows of $B'$ are again given by Eq.~\eqref{eq:big_lgs_4rows}, but with $6^{c-j}$ instead of $6^{k-j}$ blocks of the form $B_j$.
For the remaining part, we set 
the three rows from row  $(4c +3j-2)$ to row $(4c+3j)$ to
\begin{equation} 
\begin{bmatrix}
\mathbf{x}_{n-k+c+j} ^\mathrm{T}& 0 &  \cdots & 0 \\
\mathbf{w}_{n-k+c+j} ^\mathrm{T}& 1 &  \cdots & 1 \\
\mathbf{y}_{n-k+c+j} ^\mathrm{T}& 1 &  \cdots & 1 \\   
\end{bmatrix} \in \FF_2^{3\times (d+6^c)}
\end{equation}
for all $j\in\{1,\ldots, k-c\}$, {i.e.,  we dispose of the $\mathbf{z}$-row.}
In this way, we have effectively combined the two cases that correspond to $\mathbf{b}_1$ and $\mathbf{b}_2$ from Eq.~\eqref{eq:6cases}.
The remainder of our approach stays unchanged.
By restricting from $B$ to $B'$, 
we will only be able to find solutions $\boldsymbol{\lambda} \in \mathcal{L}_{n-k+1} \cap\ldots \cap \mathcal{L}_n$ which are contained in the subspace
\begin{equation}
    \label{eq:restricted_space}
    \left(\bigcap_{i=n-k+1}^{n-k+c} 
    \underbrace{\mathcal{A}_i \cup\mathcal{B}_i \cup\mathcal{C}_i \cup\mathcal{D}_i}_{=\mathcal{L}_i}\right) \cap  \bigcap_{i=n-k+c+1}^n \underbrace{\mathcal{A}_i}_{\subset \mathcal{L}_i}
    .
\end{equation}
 
Let us illustrate the working principle of the cutoff~$c$.
In the hypothetical example of Fig.~\ref{fig:geometry} in the main text, 
we have highlighted the affine subspaces 
\begin{equation}
    \mathcal{A}_i \subset \mathcal{L}_i  =  \mathcal{A}_i \cup  \mathcal{B}_i  \cup  \mathcal{C}_i  \cup  \mathcal{D}_i 
\end{equation}
with an increased line width to distinguish them from  $ \mathcal{B}_i $, $\mathcal{C}_i$, and $ \mathcal{D}_i$.
In this example, the number of rank-2 hypersurfaces is $k=3$. 
Hence, there are four possible choices for the cutoff $c\in\{0,\ldots, k\}$.
For $c=k=3$, we recover the original approach and are able to probe all six depicted intersection spaces (yellow dots). 
For $c=2$, the restricted search space $\mathcal{L}_1 \cap \mathcal{L}_2 \cap \mathcal{A}_3$ only contains two non-empty intersection spaces, namely $\mathcal{B}_1 \cap \mathcal{A}_2 \cap \mathcal{A}_3$ and $\mathcal{D}_1 \cap \mathcal{C}_2 \cap \mathcal{A}_3$.
For $c=1$, only $\diameter \neq \mathcal{B}_1 \cap \mathcal{A}_2 \cap \mathcal{A}_3 = \mathcal{L}_1 \cap \mathcal{A}_2 \cap \mathcal{A}_3$ remains.
Note that neither $\mathcal{A}_1 \cap \mathcal{A}_2$ nor  $\mathcal{A}_1 \cap \mathcal{A}_3$ nor  $\mathcal{A}_2 \cap \mathcal{A}_3$ is empty, but  $\mathcal{A}_1 \cap \mathcal{A}_2 \cap  \mathcal{A}_3$ is.
Therefore, we would not be able to find any solution for $c=0$ 
in the example of Fig.~\ref{fig:geometry}  in the main text.
For a non-hypothetical example, see {\crossrefcolor SM Sec.~XI}.

\subsection*{Runtime Analysis}

The \solver{restricted algebraic solver} consists of an outer loop over $s(n)\le 2^e$ subgraphs and an inner loop over $4^{c(n)}$ subspace intersections, where $s(n)$ and $c(n)$ are two hyperparameters that  may explicitly depend on $n$.
For each subgraph and each intersection, the  \solver{restricted algebraic solver}  applies Gaussian elimiation to a binary matrix $B'$ of size $(4c(n)+3(k-c(n))) \times (d+6^{c(n)})$, where $k\le n$ and $d\le 4n$.
If we choose a constant cutoff $c(n)=\text{const}.$~as in Eq.~\eqref{eq:cutoff}, then the size of $B'$ is proportional to $n \times n$. 
Hence, Gaussian elimination has a runtime of $\mathcal{O}(n^3)$, i.e., the \solver{restricted algebraic solver} has a runtime of $\mathcal{O}(s(n)\times n^3)$.
This is efficient if we choose $s(n)=\mathcal{O}(\poly(n))$.
In particular, for the hydrogen chain Hamiltonians in Fig.~\ref{fig:all_examples} where we work with constant values of $s(n)$, the solver should have a runtime of $f(n) = \mathcal{O}(n^3)$.
This is consistent with our empirical estimate of a runtime that lies somewhere in between $\mathcal{O}(n^{2.0})$ and $\mathcal{O}(n^{2.4})$.

On the other hand, if we choose a logarithmically-growing cutoff $c(n) = c_0  \times \log_6(n)$,
the inner loop of the \solver{restricted algebraic solver} iterates over $6^{c(n)} = n^{c_0}$ many intersections.
Furthermore, we apply Gaussian elimination of a matrix $B'$ of size proportional to $n \times n^{c_0}$, which has a runtime of $\mathcal{O}(n^{2+c_0})$ assuming $c_0\ge 1$.
Hence, the overall runtime of the \solver{restricted algebraic solver} follows as $f(n) = \mathcal{O}(s(n) \times  n^{2 c_0+2})$, which is efficient for $s(n)= \mathcal{O}(\poly(n))$.

\subsection*{Ideas for Improving our Solvers}

There is much room to further improve the runtime of both the solver for Eq.~\eqref{eq:linear_condition_A} and the Pauli grouping algorithm into which it is embedded:
\begin{enumerate}
    \item Since our solvers are highly parallelizable, a more distributed software implementation would allow us to trade runtime for computational resources.
    \item  Currently, our Pauli grouper calls the solver without making use of previously computed solutions. 
        Warm starting methods could improve upon this.
        For example, the RREF of $M$ from Eq.~\eqref{eq:LGS} could perhaps be reused.
    \item   
    %For Fig.~\ref{fig:all_examples}, we have tuned them manually. 
    Methods for automatically tuning the hyperparameters of our \solver{restricted algebraic solver} deserve further investigation.
    \item  Instead of  selecting the polynomially-large subset of subgraphs at random, one could implement a more sophisticated approach such as simulated annealing.
    \item  The geometry of the solution space $\mathcal{L}$ could be further explored.
           If we had access to a lexicographical Gröbner basis for the ideal generated by $ \boldsymbol{\lambda}^\mathrm{T} Q_1 \boldsymbol{\lambda} +1$, $\ldots$,  $ \boldsymbol{\lambda}^\mathrm{T} Q_n \boldsymbol{\lambda} +1 \in   \FF_2[\lambda_1,\ldots, \lambda_d]$ 
           (which defines the Zariski-closed set $\mathcal{L}\subset \FF_2^d$), we could construct a solution $\boldsymbol{\lambda}\in \mathcal{L}$ via elimination theory~\cite{cox_ideals_varieties_2013}.
\end{enumerate}

\section*{Data Availability}
All relevant data are available from the authors upon reasonable request.
\\ 

\section*{Code Availability}
% The underlying code for this study is not publicly available for proprietary reasons. 
An open-source C++~implementation of our Pauli grouping algorithm based on our \solver{informed numerical solver} is available under \url{https://github.com/Mc-Zen/HT-Grouper}. 
\\

\section*{Acknowledgements}
This research is part of two projects that have received funding from the European Union’s Horizon 2020 research and innovation programme under the Marie Skłodowska-Curie grant agreements No.~847471 and No.~955479.
This research was supported by the NCCR SPIN, funded by the Swiss National Science Foundation (SNF).
I.O.S. acknowledges the financial support from the SNF through the grant No.~200021-179312.
We acknowledge the use of IBM Quantum services for this work.
The views expressed are those of the authors, and do not reflect the official policy or position of IBM or the IBM Quantum team.

In the final phase of this project, DM received financial support from the Munich Quantum Valley (K-8), the Bundesministerium für Bildung und Forschung (BMBF) via the projects HYBRID, REALISTIQ, and MUNIQC-Atoms as well as from the EU Quantum Technology Flagship via the project MILLENION.
Research was sponsored by IARPA and the Army Research Office, under the Entangled Logical Qubits program, and was accomplished under Cooperative Agreement Number  W911NF-23-2-0212. The views and conclusions contained in this document are those of the authors and should not be interpreted as representing the official policies, either expressed or implied, of IARPA, the Army Research Office, or the U.S. Government. The U.S. Government is authorized to reproduce and distribute reprints for Government purposes notwithstanding any copyright notation herein. 

The authors are thankful to Lennart Vincent Bittel, Libor Caha, Felix Huber, Seyed Sajjad Nezhadi, Max Rossmannek, Maria Spethmann, David Sutter, Stefan Woerner, James Robin Wootton, and Nikolai Wyderka for stimulating discussions.
\hfill
SDG
\\

\section*{Competing Interests}
The authors declare no competing financial or non-financial interests.
\\

\section*{Author Contributions}
DM developed the theory, implemented and tested the grouping algorithm, and wrote the manuscript.
DM and PB performed the experiment. DM and LF analyzed the experimental data.
DM, LF, IS, PB, and IT were involved in the design of the experiment and discussed the results.
KL implemented the open-source version of the grouping algorithm.
KL, EK, JE, and IS supported DM in carrying 
out the study in SM~Sec.~XIV.
All authors contributed to the manuscript.\\

% Declarations

\providecommand{\noopsort}[1]{}\providecommand{\singleletter}[1]{#1}%

\end{document}